\documentclass[conference]{IEEEtran}
\IEEEoverridecommandlockouts
\usepackage{amsmath,amsfonts}
\usepackage{algorithmic}
\usepackage{algorithm}
\usepackage{array}
\usepackage[caption=false,font=normalsize,labelfont=sf,textfont=sf]{subfig}
\usepackage{textcomp}
\usepackage{stfloats}
\usepackage{url}
\usepackage{verbatim}
\usepackage{graphicx}
\usepackage{caption}
\usepackage{romannum}
\usepackage{cite}
\usepackage{amsmath}
\usepackage{makecell}
\usepackage{multirow}
\usepackage{hyperref}
\usepackage{xcolor}

\begin{document}
\vspace{-6pt}
\title{\fontsize{22}{24}\selectfont Physics-Aware LLM-Based Probabilistic Wind Power Scenario Generation under Extreme Icing Conditions
\thanks{This research was supported by the National Science Foundation through award numbers 2418359 and 2346213.}%
}
\vspace{-6pt}
\author{\fontsize{16}{18}\selectfont 
\IEEEauthorblockN{ Lei Wang,Ying Zhang}
\IEEEauthorblockA{\textit{\makecell[c]{School of Electrical and Computer \\[-1pt]Engineering}} \\[-1pt]
\textit{Oklahoma State University},\\[-1pt]
Stillwater, OK, U.S. \\[-1pt]
\{leiwang;y.zhang\}@okstate.edu}
\and
\IEEEauthorblockN{Di Shi}
\IEEEauthorblockA{
\textit{\makecell[c]{Klipsch School of Electrical and Computer \\ [-1pt]
Engineering}} \\[-1pt]
\textit{New Mexico State University},\\[-1pt]
Las Cruces, NM, U.S. \\[-1pt]
dshi@nmsu.edu}
\and
\IEEEauthorblockN{Fei Ding}
\IEEEauthorblockA{
\textit{National Renewable Energy Laboratory} \\[-1pt]
Golden, CO, U.S. \\[-1pt]
fei.ding@nrel.gov}
} 
\maketitle
\vspace{-100pt}
\begin{abstract}
Accurately characterizing wind power uncertainty under icing and post-disaster conditions remains a critical challenge for resilient power system operation. To address this issue, this paper proposes a physics-aware large language model (LLM) framework for probabilistic wind power scenario generation under extreme icing conditions. The proposed framework integrates supervisory control and data acquisition (SCADA)-based physical modeling, multimodal tokenization, and a causal Transformer architecture trained in an autoregressive manner. A physics-aware decoding scheme effectively enforces rated power limits and ramping constraints on the generated trajectories while preserving stochastic diversity. Case studies using real wind turbine data show that the proposed method reproduces icing-induced power degradation and temporal variability observed during extreme weather. The resulting scenarios are physically consistent and high-fidelity, thereby significantly enhancing resilience assessment and recovery planning in renewable-integrated power systems.
\end{abstract}


\section{Introduction} 
\IEEEPARstart{E}{xtreme} weather events such as icing, blizzards, and storms increasingly threaten the reliable operation of renewable-integrated power systems. In cold-climate regions, wind farms are particularly vulnerable to blade icing, which causes aerodynamic degradation, mechanical stress, and severe power losses \cite{wang2024novel}. When such events occur on a large scale, the resulting generation uncertainty propagates across the grid’s temporal and spatial scales, leading to voltage instability, load shedding, and delayed system recovery \cite{athari2017impacts}. Following such disasters, resilient operations and restoration rely critically on accurate modeling of renewable generation uncertainty \cite{xu2024resilience}. Hence, realistic post-disaster wind power scenarios provide a foundation for resilience assessment, restoration optimization, and secure operation planning in renewable-integrated power systems \cite{li2020review}.

Existing scenario generation approaches can be broadly classified into explicit and implicit data-driven approaches. Explicit methods rely on predefined probabilistic models to characterize renewable generation uncertainty \cite{rayati2023stochastic,krishna2023time,chen2012probabilistic}. Although interpretable and computationally efficient, these methods usually assume linearity, stationarity, or fixed distributions, limiting their ability to capture nonlinear and non-stationary wind power characteristics under extreme conditions.

In contrast, implicit methods directly learn the probability distribution of wind power from historical data without relying on explicit parametric assumptions \cite{chen2018model,yuan2022conditional,li2024novel,dong2025controllable}. Despite their strong expressive power, most existing data-driven methods are designed for normal operating conditions and neglect the physical impacts of disasters such as icing. As a result, generated scenarios may violate physical constraints or deviate from realistic post-disaster behaviors.

Despite these advances, most data-driven methods are designed for normal operating conditions and neglect the physical impacts of disasters such as icing or storms. These events alter turbine aerodynamics and mechanical behavior, introducing nonlinear and degradation-driven dynamics that purely statistical models fail to represent. As a result, the 
generated scenarios may violate physical constraints or deviate from realistic post-disaster behaviors, reducing their value for resilience-oriented applications.

To address these limitations, there is a growing need for physics-aware, data-driven frameworks that integrate physical knowledge with learning-based generative models. Recent developments in large language models (LLMs) provide a promising foundation, offering strong capabilities in temporal reasoning and probabilistic sequence generation. Compared with variational autoencoder (VAE)-, conditional generative adversarial network (CGAN)-, or diffusion-based generators, an autoregressive LLM models wind power as a causal sequence, which better matches long-horizon, non-stationary icing dynamics. Its stepwise decoding also makes it straightforward to impose operational limits (e.g., rated-power and ramping constraints) during sampling without sacrificing diversity. By combining LLM-based generative modeling with physics-informed constraints, this study aims to produce realistic, interpretable, and physically consistent wind power scenarios under disaster conditions. The main contributions of this study are summarized as follows:

\begin{itemize}
\item A physics-aware LLM framework is proposed to 
generate probabilistic wind power scenarios under icing conditions. By integrating physical priors—such as rated power limits, ramping constraints, and icing-induced degradation behaviors—into the generative process, the framework ensures 
the physical consistency of the generated trajectories, enabling reliable use for resilience analysis and restoration planning.
\item A causal Transformer architecture with multimodal tokenization is designed to capture nonlinear, stochastic, and context-dependent features in wind power data. This mechanism 
enhances the interpretability and reliability of scenario generation, outperforming conventional data-driven approaches such as VAE-, CGAN-, and diffusion-based models in modeling uncertainty under extreme weather conditions.
\end{itemize}
\section{Problem Formulation}
Let $x_t$ denote the wind power output at time step $t$. The 
historical wind power sequence of length $L$ is represented as 
$x_{t-L+1:t}=\{x_{t-L+1},\dots,x_t\}$. At each time step, we denote the 
exogenous feature vector as $\mathbf{f}_t$, which includes supervisory 
control and data acquisition (SCADA) measurements (e.g., pitch 
angle, rotor speed, nacelle internal temperature) and 
meteorological variables (e.g., wind speed, ambient 
temperature). Accordingly, $\mathbf{f}_{t-L+1:t}$ denotes the historical feature 
sequence aligned with $x_{t-L+1:t}$.

In this work, we focus exclusively on icing disaster scenarios, 
where the historical wind power and feature sequences are 
collected under icing conditions. This ensures that the model 
explicitly learns the temporal dynamics of wind power 
degradation caused by blade icing.

Given the historical power sequence $x_{t-L+1:t}$ and the 
associated historical features $\mathbf{f}_{t-L+1:t}$, the task of probabilistic 
scenario generation is to produce a set of $M$ plausible future 
trajectories of wind power:
\begin{equation}
\left\{\tilde{x}^{(m)}_{t+1:t+H}\right\}_{m=1}^{M} \sim 
P\!\left(x_{t+1:t+H}\mid x_{t-L+1:t},\,\mathbf{f}_{t-L+1:t}\right),
\end{equation}
where
$\tilde{x}^{(m)}_{t+1:t+H}
=
\left\{
\tilde{x}^{(m)}_{t+1},
\tilde{x}^{(m)}_{t+2},
\dots,
\tilde{x}^{(m)}_{t+H}
\right\}$
denotes the $m$-th generated scenario. By concentrating on icing-
affected data, the model captures the degradation patterns and 
enhanced variability induced by icing. The generated scenarios 
are post-processed with turbine physical limits (e.g., rated power 
cap and ramping constraints) to ensure physical plausibility.
\vspace{-2pt}
\section{Proposed Physics-Aware LLM Method}
\begin{figure}[!t]
    \centering
        \includegraphics[width=0.45\textwidth]{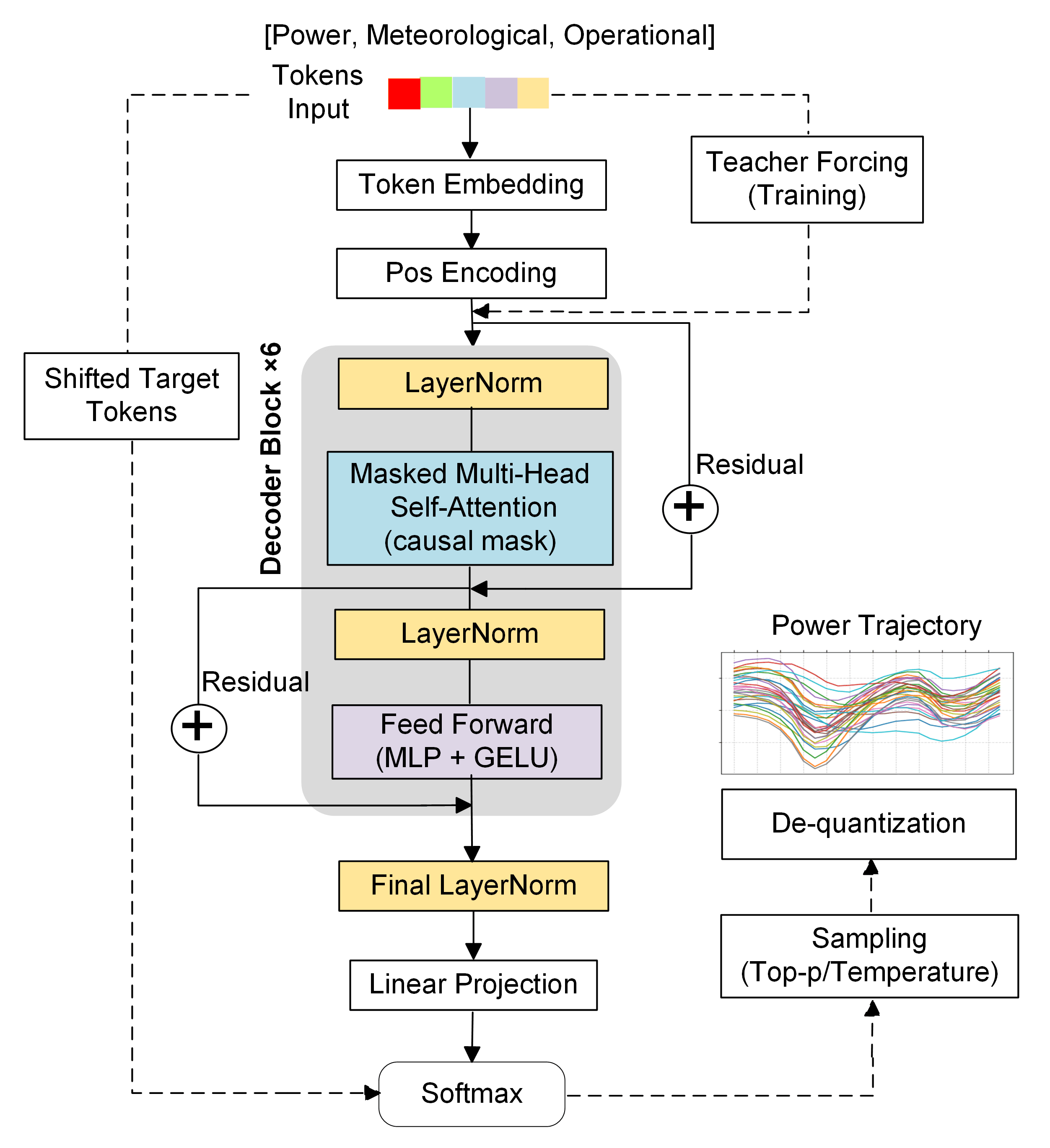}
    \caption{ LLM-based Causal Transformer architecture for wind power 
scenario generation under icing conditions.}
    \label{fig:1}
    \vspace{-12pt}
\end{figure}

The proposed framework, illustrated in Fig. \ref{fig:1}, generates 
probabilistic wind power scenarios under icing conditions by 
capturing temporal dependencies in power dynamics and 
enforcing physics-based constraints. It processes historical wind 
power, SCADA, and meteorological data, with the feature set 
optimized through correlation analysis with the target output. 
The workflow includes: i) Feature Selection: Retaining SCADA 
and meteorological variables strongly correlated with wind 
power to reduce noise and improve efficiency; ii) Physical 
Model Identification: Fitting a wind speed–power curve to non-icing data via least squares to establish a normal-operation 
baseline that constrains subsequent generation \cite{wang2024wind}; iii) 
Tokenization: Discretizing the selected features into tokens 
according to their physical ranges and concatenating them into 
multimodal input sequences; iv) LLM-based Modeling: Feeding 
tokenized sequences into a decoder-only causal Transformer 
trained autoregressively with teacher forcing to predict future 
power tokens; v) Physics-aware Training and Decoding:
Incorporating physics-based penalties—such as rated power 
caps, baseline bounds, ramp-rate limits, and smoothness—while 
applying constrained sampling to enforce physical feasibility; 
and vi) Scenario Generation: De-quantizing sampled tokens 
into continuous trajectories and producing multiple scenarios via 
top-p and temperature sampling that reflect icing-induced 
degradation with a lower mean and higher variability.

\subsection{Tokenization and LLM Architecture}

The goal of tokenization is to transform continuous turbine measurements into discrete tokens suitable for autoregressive 
modeling with LLMs \cite{zhang2024large}. In this study, we specifically focus on icing scenarios, and therefore, only on turbine operation under iced conditions is considered. Accordingly, SCADA and 
meteorological features are normalized and discretized into 
fixed bins, as summarized in Table I.

Each feature is first scaled to the interval [0, 1] according to 
its physical or operational range. The wind power output, 
normalized by rated capacity, is quantized into 256 $\mu$-law bins 
to emphasize detail in the low-production regime while 
compressing the high-power region. Meteorological variables 
are discretized at different resolutions: wind speed into 64 bins, 
and temperature into 16 bins. Operational variables, such as 
pitch angle, yaw position, and generator speed, are quantized 
into 16 bins to encode control dynamics under icing conditions.

At every time step, the tokenized features are concatenated 
into a multimodal tuple [Power, Meteorological, Operational]. 
The temporal sequence of such tuples then forms the model 
input, enabling the Transformer to autoregressively generate 
wind power trajectories that reflect both environmental and 
operational contexts.

At time step $t$, the tokenized input sequence is denoted as
\begin{equation}
\mathbf{x}_{t}=\{x^{(p)}_{t},\,x^{(m)}_{t},\,x^{(o)}_{t}\},\quad t=1,\dots,L,
\end{equation}
where $x^{(p)}_{t}$, $x^{(m)}_{t}$, and $x^{(o)}_{t}$ represent power, meteorological, and 
operational tokens, respectively. Each discrete token index is 
mapped into a dense representation by a learnable embedding 
matrix $\mathbf{E}\in\mathbb{R}^{V\times d}$, with $V$ the vocabulary size and $d$ the 
embedding dimension. Positional encodings $\mathbf{p}_{t,k}\in\mathbb{R}^{d}$ are 
added to preserve temporal order, yielding
\begin{equation}
\mathbf{z}_{t,k}=\mathbf{E}[x^{(k)}_{t}]+\mathbf{p}_{t,k},
\end{equation}
where $k \in \{p,m,o\}$ indexes the modalities corresponding to 
power, meteorological, and operational tokens, respectively.

\begin{table}[!t]
\caption{Tokenization Settings for Features}
\label{tab:token}
\centering
\begin{tabular}{|l|l|l|}
\hline
Feature type & Variables & Discretization (bins)\\
\hline
Power & Active power & $\mu$-law 256 ($\mu$ = 120)\\
\hline
Meteorological & Wind speed, temp. & Quantile 64 (wind), 16 (temp.)\\
\hline
Operational & \makecell[l]{Pitch angle, yaw, \\generator speed} & Quantile 16\\
\hline
\end{tabular}
\vspace{-5 pt}
\end{table}

Concatenating all modalities across the temporal dimension 
gives the model input sequence
\begin{equation}
\mathbf{Z}^{(0)}=\{z_{1,p},\,z_{1,m},\,z_{1,o},\,\dots,\,z_{L,p},\,z_{L,m},\,z_{L,o}\}.
\end{equation}

The embedded sequence is passed through a stack of $N$ 
decoder blocks, each following a design where layer 
normalization is applied before every sub-layer and residual 
connections are preserved. Given the input representation 
$\mathbf{H}^{(l-1)}\in\mathbb{R}^{L\times d}$ to the $l$-th block, masked multi-head self-attention 
is first applied. For each attention head $h =1,\dots, H$, the query, 
key, and value matrices are computed as
\begin{equation}
\mathbf{Q}_{h}=\mathbf{X}\mathbf{W}^{Q}_{h},\ \ \mathbf{K}_{h}=\mathbf{X}\mathbf{W}^{K}_{h},\ \ \mathbf{V}_{h}=\mathbf{X}\mathbf{W}^{V}_{h},
\end{equation}
where $\mathbf{X}=\mathrm{LN}(\mathbf{H}^{(l-1)})$, $\mathbf{W}^{Q}_{h},\mathbf{W}^{K}_{h},\mathbf{W}^{V}_{h}\in\mathbb{R}^{d\times d_{k}}$ are learnable 
projection matrices, $d$ is the embedding dimension, and $d_{k}=d/H$ 
is the dimension per head. A causal mask $\mathbf{M}_{\mathrm{causal}}$ is applied to 
prevent attending to future tokens. The attention output for head 
$h$ is
\begin{equation}
\mathrm{Attn}_{h}=\mathrm{Softmax}\!\left(\frac{\mathbf{Q}_{h}\mathbf{K}_{h}}{\sqrt{d_{k}}}+\mathbf{M}_{\mathrm{causal}}\right)\mathbf{V}_{h}.
\end{equation}

Outputs from all heads are concatenated and linearly 
projected, followed by a residual connection:
\begin{equation}
\mathbf{U}^{(l)}=\mathbf{H}^{(l-1)}+\mathrm{Concat}(\mathrm{Attn}_{1},\dots,\mathrm{Attn}_{H})\mathbf{W}^{O},
\end{equation}
where $\mathbf{W}^{O}\in\mathbb{R}^{Hd_{k}\times d}$ is the output projection.

The second sub-layer is a position-wise feed-forward 
network with GELU activation,
\begin{equation}
\mathrm{FFN}(\mathbf{x})=\mathbf{W}_{2}\,\sigma(\mathbf{W}_{1}\mathbf{x}+\mathbf{b}_{1})+\mathbf{b}_{2},
\end{equation}
where $\mathbf{W}_{1}\in\mathbb{R}^{d\times d_{\mathrm{ff}}}$, $\mathbf{W}_{2}\in\mathbb{R}^{d_{\mathrm{ff}}\times d}$ are learnable weights, $\mathbf{b}_{1},\mathbf{b}_{2}$ are 
biases, and $d_{\mathrm{ff}}$ is the hidden dimension of the feed-forward layer. 
After applying layer normalization, the output is added through 
another residual connection, yielding the block output
\begin{equation}
\mathbf{H}^{(l)}=\mathbf{U}^{(l)}+\mathrm{FFN}\!\left(\mathrm{LN}(\mathbf{U}^{(l)})\right).
\end{equation}

After $N$ stacked decoder blocks, the final hidden states are 
normalized and projected into the vocabulary space:
\begin{equation}
\mathbf{o}_{t}=\mathrm{Softmax}\!\left(\mathbf{W}_{o}\,\mathrm{LN}(\mathbf{h}^{(N)}_{t})\right),\quad \mathbf{o}_{t}\in\mathbb{R}^{V}
\end{equation}
where $\mathbf{h}^{(N)}_{t}\in\mathbb{R}^{d}$ is the hidden representation of the $t$-th token 
after the last block, $\mathbf{W}_{o}\in\mathbb{R}^{V\times d}$ is the output projection matrix, 
and $V$ is the vocabulary size determined by the discretization 
bins across all features. The output vector $\mathbf{o}_{t}$ is a probability 
distribution over the next possible tokens.

During training, the model is optimized with teacher forcing, 
where the shifted ground-truth sequence $\{x_{2},\dots,x_{L+1}\}$ serves as 
the prediction target. The training objective is the cross-entropy 
loss
\begin{equation}
\mathcal{L}_{\mathrm{CE}}=-\sum_{t=1}^{L}\log \mathbf{o}_{t}[x_{t+1}],
\end{equation}
where $\mathbf{o}_{t}[x_{t+1}]$ denotes the predicted probability assigned to the 
true next token $x_{t+1}$.

During inference, tokens are generated autoregressively as
\begin{equation}
x_{t+1}\sim \mathbf{o}_{t},
\end{equation}
using nucleus sampling with top-$p$ filtering and temperature 
scaling. The resulting discrete sequence is finally de-quantized 
back into continuous wind power trajectories.

\subsection{Physics-aware Scenario Generation}

After autoregressive modeling in Section III-A, the model 
outputs a probability distribution $\mathbf{o}_{t}\in\mathbb{R}^{V}$ over possible next 
power tokens at each time step $t$. Let $q(\cdot)$ and $q^{-1}(\cdot)$ denote the 
quantization and de-quantization mappings, respectively, with 
$q^{-1}:\{1,\dots,V_{P}\}\to[0,1]$ mapping discrete tokens back to 
normalized power values. The generated token $\hat{x}_{t+1}$ is decoded 
into power as
\begin{equation}
\hat{p}_{t+1}=q^{-1}(\hat{x}_{t+1})\cdot P_{\mathrm{rated}},
\end{equation}
where $P_{\mathrm{rated}}$ is the rated capacity of the turbine. Meteorological 
and operational features are not predicted, but instead serve as 
conditioning variables to guide the generation process, ensuring 
that power trajectories reflect the actual operating environment 
under icing conditions.

To enforce physical plausibility, three classes of constraints 
are imposed. First, a power cap loss prevents predicted outputs 
from exceeding either the rated power or a scaled physical 
baseline $P_{\mathrm{norm}}(v_{t})$, obtained from our previous work \cite{wang2024wind}:
\begin{equation}
\begin{aligned}
\mathcal{L}_{\mathrm{cap}}
&=
\frac{1}{L}
\sum_{t=1}^{L}
\Big[
\hat{p}_{t}
-
\min\!\big(
P_{\mathrm{rated}},
\alpha P_{\mathrm{norm}}(v_{t})
\big)
\Big]_{+},
\\
[z]_{+} &= \max(z, 0),
\end{aligned}
\label{eq:Lcap_and_relu}
\end{equation}
where $0< \alpha \le 1$ is a safety margin factor.

Second, a ramping loss penalizes sharp variations between 
consecutive steps:
\begin{equation}
\mathcal{L}_{\mathrm{ramp}}=\frac{1}{L}\sum_{t=2}^{L}\Big[\ (\hat{p}_{t+1}-\hat{p}_{t}-R^{\uparrow})_+^{2}+(\hat{p}_{t}-\hat{p}_{t+1}-R^{\downarrow})_+^{2}\ \Big]
\end{equation}
where $R^{\uparrow}$ and $R^{\downarrow}$ are upward and downward ramp-rate limits 
derived from the historical data.

Third, a temporal smoothness penalty is included via a 
Huber-style total variation term:
\begin{equation}
\mathcal{L}_{\mathrm{tv}}=\frac{1}{L}\sum_{t=2}^{L}\rho_{\delta}(\hat{p}_{t}-\hat{p}_{t-1}),
\end{equation}
where $\rho_{\delta}$ denotes the Huber loss.

The overall objective combines data likelihood with physics-
based regularization:
\begin{equation}
\mathcal{L}=\mathcal{L}_{\mathrm{CE}}+\lambda_{\mathrm{cap}}\mathcal{L}_{\mathrm{cap}}+\lambda_{\mathrm{ramp}}\mathcal{L}_{\mathrm{ramp}}+\lambda_{\mathrm{tv}}\mathcal{L}_{\mathrm{tv}}
\end{equation}
where $\lambda_{\mathrm{cap}}$, $\lambda_{\mathrm{ramp}}$, and $\lambda_{\mathrm{tv}}$ are balancing hyperparameters.

During inference, physics-aware sampling is applied. At 
each step $t$, candidate power tokens are first drawn using nucleus 
sampling with temperature scaling. The candidate set $\mathcal{C}_{t}$ is then 
pruned by discarding those that violate cap or ramping limits:
\begin{equation}
\begin{aligned}
\mathcal{C}_{t} \leftarrow 
\Big\{\, i \in \mathcal{C}_{t} \ \big| \ 
& p(i) \le \min\!\big(P_{\mathrm{rated}},\, \alpha P_{\mathrm{norm}}(v_{t})\big), \\
& \big|p(i)-\hat{p}_{t}\big| \le \tilde R
\Big\}
\end{aligned}
\label{eq:candidate_set}
\end{equation}
where $\tilde{R}$ is a relaxed ramping tolerance. If no feasible candidate 
remains, the nearest valid token is projected back. Repeating this 
process over $H$ steps produces a complete power trajectory.

Finally, the discrete sequence $\{\hat{x}_{L+1},\dots,\hat{x}_{L+H}\}$ is de-
quantized into continuous values $\{\hat{p}_{L+1},\dots,\hat{p}_{L+H}\}$. Light 
smoothing (e.g., moving average within physical limits) is 
optionally applied to suppress spurious spikes. Sampling 
multiple trajectories yields a diverse yet physically consistent 
scenario set $\{\hat{p}^{(m)}_{L+1:L+H}\}_{m=1}^{M}$, aligning probabilistic generation 
with both data-driven patterns and turbine physics.
\section{Case Study}
The proposed LLM-based 
stochastic scenario generation framework is implemented in 
Python 3.8.2 using the PyTorch 1.8.0 GPU backend. All experiments are conducted on a workstation equipped 
with an Intel Xeon E5-2680 v4 CPU (2.4 GHz) and an NVIDIA 
GeForce RTX 3080 Ti GPU. 

The SCADA dataset from two wind turbines in North China ~\cite{wang2024wind} is adopted. Following standard preprocessing in~\cite{wang2024wind}, the data is cleaned and resampled to 1-min resolution, and 26 SCADA variables are retained and normalized to $[0,1]$. Non-icing periods are used to fit a four-parameter physical power curve and estimate rated power and ramping limits, while icing periods are used for training and evaluation. The data are split chronologically into 70\%/15\%/15\% for training/validation/testing.

The model performance is evaluated using three criteria: (i) the 
continuous ranked probability score (CRPS)~\cite{gilbert2019leveraging} for
probabilistic accuracy, (ii) the Kullback--Leibler divergence 
(KLD)~\cite{dong2025controllable} for statistical similarity, and (iii) the violation 
rate against physical constraints (rated power and ramping 
limits) for physical consistency.
\subsection{Tokenization \& Scenario Generation}
\begin{figure}[!t]
    \centering
        \includegraphics[width=0.5\textwidth]{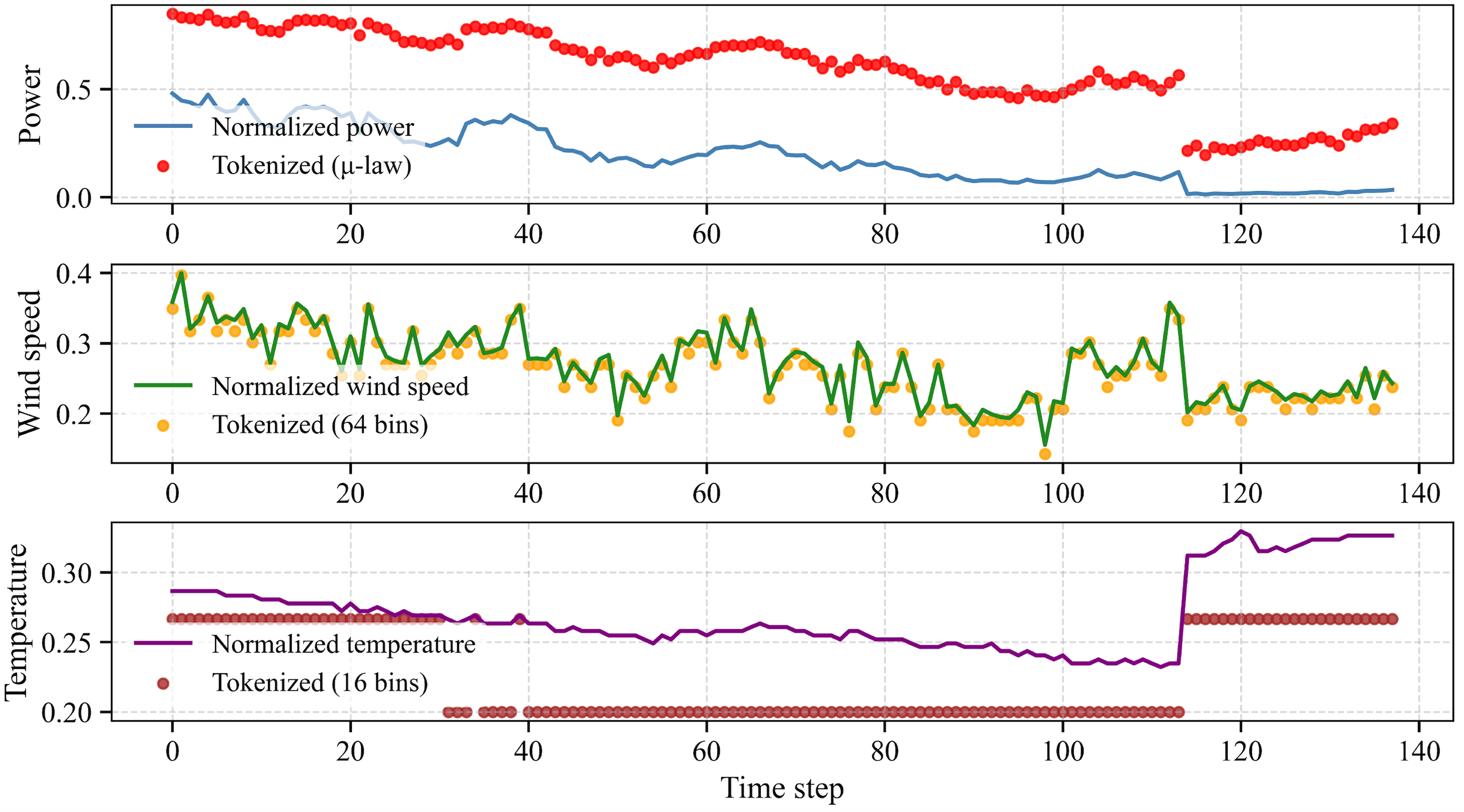}
    \caption{Tokenization results for power, wind speed, and temperature.}
    \label{fig:2}
    \vspace{-15pt}
\end{figure}

To enable the LLM-based scenario generation, continuous 
SCADA features are transformed into discrete tokens. Each 
feature is normalized using the statistics of non-icing data to 
ensure consistent scaling between normal and icing conditions. 
Then, the normalized signals are discretized according to their 
physical or operational ranges, as summarized in Table~\ref{tab:token}. 
The active power is quantized into 256 $\mu$-law bins to emphasize 
variations in low-output regions, while wind speed and 
temperature are uniformly discretized into 64 and 16 bins, 
respectively. This feature-specific quantization preserves the 
physical dynamics of different variables while reducing 
dimensional complexity. The tokenized sequences of power, 
wind speed, and temperature (Fig. \ref{fig:2}) clearly reflect the 
underlying temporal patterns and serve as structured symbolic 
inputs for the subsequent LLM-based scenario generation.

\begin{figure}[!t]
\centering
\subfloat[]{%
\includegraphics[width=4.3cm]{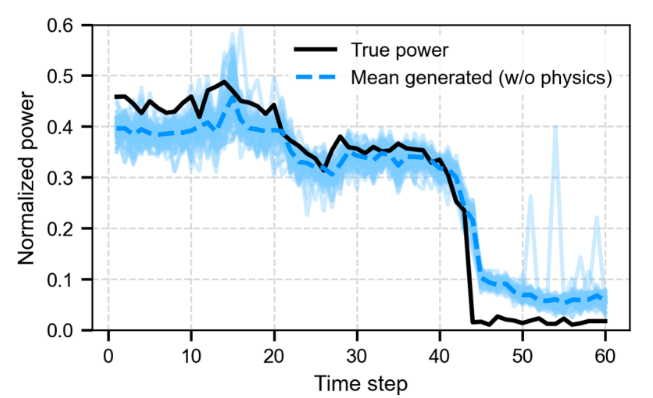}
}
\hfill
\subfloat[]{%
\includegraphics[width=4.3cm]{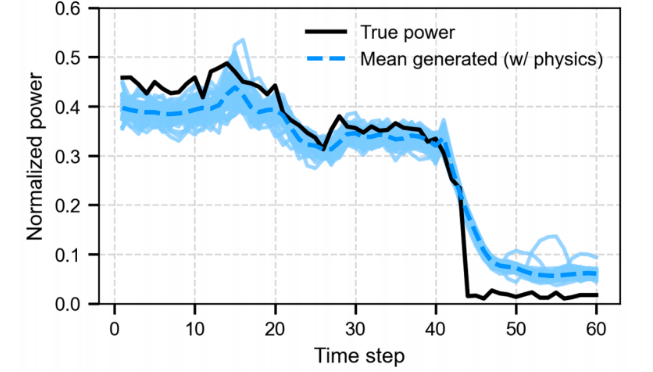}
}
\vspace{-5pt}
\caption{Comparison of LLM-generated wind power scenarios without and with physics-aware constraints.}
\label{fig:3}
\vspace{-5pt}
\end{figure}

To evaluate the stochastic scenario generation capability and 
the impact of physics-aware constraints, 50 wind power 
trajectories are generated conditioned on the predicted 
normalized power. Fig.\ref{fig:3} (a) shows the results without physical 
constraints, where the black solid line denotes the true power, 
the light blue curves represent unconstrained LLM-generated 
scenarios, and the blue dashed line shows their mean. The 
unconstrained results accurately capture the overall temporal 
evolution and nonlinear variations of the true curve, including 
the abrupt drop around step 40, but exhibit occasional unrealistic 
spikes and excessive fluctuations due to the absence of physical 
regularization. In contrast, Fig.\ref{fig:3}(b) presents the results with 
physics-aware constraints including the power-cap loss, ramp-rate 
penalty, and temporal smoothness regularization. These 
constraints effectively suppress non-physical excursions while 
preserving the stochastic diversity of the generated samples. The 
trajectories exhibit improved temporal consistency and physical 
plausibility, particularly in the post-event region, confirming 
that the proposed physics-aware sampling enforces realistic 
operational dynamics without degrading probabilistic variability. 
Overall, these results demonstrate that the proposed LLM-based 
framework can serve as a powerful probabilistic yet physically 
consistent generator for uncertainty quantification and resilient 
operation analysis in power systems.
\subsection{Performance Benchmarking and Baseline Comparison}
\begin{table}[!t]
\centering
\caption{Performance Comparison of Different Scenario Generation Models}
\label{tab:performance}
\begin{tabular}{|l|l|l|l|}
\hline
Model & CRPS & KLD & VR (\%) \\
\hline
CGAN~\cite{yuan2022conditional} & 0.162 & 0.061 & 7.05 \\
\hline
VAE~\cite{li2024novel}& 0.189 & 0.076 & 7.72 \\
\hline
Diffusion~\cite{dong2025controllable} & 0.145 & 0.052 & 6.58 \\
\hline
\textbf{Proposed LLM} & \textbf{0.121} & \textbf{0.038} & \textbf{2.85} \\
\hline
\end{tabular}
\vspace{-6pt}
\end{table}

To evaluate the effectiveness of the proposed LLM-based 
scenario generator, we compare it with three representative 
deep generative baselines, including the CGAN\cite{yuan2022conditional}-, VAE~\cite{li2024novel}-, and diffusion~\cite{dong2025controllable}-based methods. All models are assessed by the CRPS, KLD, and violation rate of the rated-power and ramping constraints.
As summarized in Table~\ref{tab:performance}, the proposed 
LLM-based method achieves the lowest CRPS and KLD among other methods, 
indicating improved probabilistic fidelity and better alignment with the statistical characteristics of ground-truth wind power trajectories. Moreover, its violation rate is significantly lower (2.85\%), demonstrating enhanced compliance with physical 
limits. These results collectively show that the LLM captures nonlinear temporal patterns more effectively and produces diverse yet physically credible scenarios, outperforming other generative approaches.

To assess if the projection-back fallback in \eqref{eq:candidate_set} reduces stochastic diversity during volatile icing events, a decoding sensitivity study without retraining is performed. We select high-volatility icing windows from the test set and generate 50 scenarios per window under three settings: Unconstrained, Default (paper), and Stricter (tighter ramp/cap). The overall violation rate (same as Table \ref{tab:performance}), a compact diversity score (average cross-scenario variability over the horizon), and the projection rate (fraction of steps triggering projection-back) are reported in Table \ref{tab: Sensitivity}. It can be shown that Default achieves a low violation rate with non-trivial diversity and a low projection rate, while stricter constraints further reduce the violation rate at the cost of reduced diversity and more frequent fallback.

\begin{table}[t]
\caption{Decoding Sensitivity on Volatile Icing Windows}
\label{tab: Sensitivity}
\centering
\setlength{\tabcolsep}{7pt}
\renewcommand{\arraystretch}{1.05}
\begin{tabular}{lccc}
\hline
Setting & VR$_{\text{total}}$ (\%) $\downarrow$ & Diversity $\uparrow$ & Projection (\%) $\downarrow$ \\
\hline
Unconstrained & 7.0 & 0.092 & 0.0 \\
Default & 3.3 & 0.079 & 1.2 \\
Stricter & 2.4 & 0.066 & 3.8 \\
\hline
\end{tabular}
 \vspace{-5pt}
\end{table}

\section{Conclusions}
This paper proposes a physics-aware LLM framework for probabilistic wind power scenario generation under icing conditions. The proposed method captures nonlinear temporal dependencies and icing-induced degradation while maintaining physical consistency through physics-aware decoding. Case studies demonstrate superior probabilistic accuracy, statistical consistency, and physical realism compared with the VAE,  CGAN, and diffusion-based baselines. Beyond producing high-fidelity probabilistic scenarios, the framework provides a generative foundation for post-disaster 
resilience analysis, supporting data-driven evaluation of 
recovery strategies, uncertainty-aware restoration optimization, 
and resilient operation planning under extreme weather.
\IEEEpubidadjcol

\bibliographystyle{IEEEtran}
\bibliography{Citation}
\let\mybibitem\bibitem

\end{document}